\documentclass[aps, pra, twocolumn, groupedaddress, nofootinbib]{revtex4-1}

\usepackage{graphicx}
\usepackage{amsmath}
\usepackage{epsfig}
\usepackage{dcolumn}
\usepackage{bm}
\usepackage{color}
\usepackage{epstopdf}

\newcommand{\nc}{\newcommand}
\nc{\eq}{\begin{equation}}
\nc{\eeq}{\end{equation}}
\nc{\eqa}{\begin{eqnarray}}
\nc{\eeqa}{\end{eqnarray}}
\nc{\ar}{\begin{array}}
\nc{\ear}{\end{array}}
\nc{\bfig}{\begin{figure}}
\nc{\efig}{\end{figure}}
\nc{\dg}{\dagger}
\nc{\sx}{\sigma_x}
\nc{\sy}{\sigma_y}
\nc{\sz}{\sigma_z}
\nc{\spl}{\sigma_+}
\nc{\sm}{\sigma_-}
\nc{\nn}{\nonumber}
\nc{\noi}{\noindent}
\nc{\adg}{a^{\dg}}
\nc{\kvec}{\mathbf{k}}
\nc{\xvec}{\mathbf{x}}

\def\bra#1{\mathinner{\langle{#1}|}}
\def\ket#1{\mathinner{|{#1}\rangle}}

\begin{document}

\title{Dissipative Landau-Zener level crossing subject to continuous measurement:\\Excitation despite decay}

\author{P. Haikka}
 \email{pinja@phys.au.dk}
\author{K. M\o lmer}%
 \email{moelmer@phys.au.dk}
\affiliation{Department of Physics and Astronomy, Aarhus University, Ny Munkegade 120, DK-8000 Aarhus C, Denmark}

\date{\today}
             
\begin{abstract}
The Landau-Zener formula provides an analytical expression for the final excitation of a quantum system after passage of an avoided crossing of two energy levels. If the two levels correspond to a ground state, and to an excited state which is subject to radiative decay, the probability of exciting the system by adiabatic passage of the level crossing is reduced. In this article we use a stochastic master equation to study the level crossing dynamics when the system is subject to continuous probing of the emitted radiation. The measurement backaction on the system associated with the fluctuating homodyne detection record alters the level crossing dynamics, leading to significant excitation in spite of decay and imperfect transfer.
\end{abstract}

\maketitle

\section{Introduction}

Two-level quantum systems driven across avoided crossings appear in a variety of physical contexts, from atomic and molecular systems \cite{nikitin}, to solid state devices with artificial atoms \cite{superconducting}, and even effective models of nonequilibrium phase transitions \cite{kibble}. Outside the strictly adiabatic regime, the probability of successfully transferring population from one quantum state to another during such sweep was first studied in 1932 by several physicists: Landau \cite{landau}, Zener \cite{zener}, Stueckelberg \cite{stueckelberg} and Majorana \cite{majorana} all solved what is now commonly known as the Landau-Zener (LZ) problem. Besides fundamental interest, adiabatic transitions provide a valuable experimental method for exciting quantum systems in a reliable and controlled way \cite{exp}. They can also be used as an interferometric tool \cite{interferometry}, and as a spectroscopic device for resolving the energy-level structure of atoms \cite{spectroscopy}.

While the original formulation of the LZ transition probability assumes a unitarily evolving quantum state vector, more recent studies have included the effect of dissipation and environmental noise on the LZ problem (see, for example, \cite{kayanuma, ao, hanngi, thorwart}). In several scenarios where dephasing is the only form of dissipation, the transition probability towards a final excited state has been found to be unaltered by the environment. In the case of radiative decay, instead, the final excited state probability is reduced below the dissipationless value \cite{hanngi}.

In this article we focus on the latter case and investigate what happens when the radiation emitted during the decay of a single quantum system is being monitored. This is a realistic situation with laser driven two-level atoms that decay by spontaneous emission of light, and it also pertains to experiments on superconducting qubits \cite{siddiqi}. We assume the radiation to be detected by a homodyne measurement scheme, where the information inscribed over time in the measurement current results in a continuous, diffusive backaction on the quantum state of the emitter \cite{wiseman, gardiner1, gardiner2}. The measurement affects the dynamics of the emitter in an unpredictable way, but based on stochastic simulations we identify a large number of trajectories where the backaction drives the system to highly excited states in the presence of the dissipative noise. For near-adiabatic sweeps it turns out to be a typical feature for the quantum trajectories to attain atypical degrees of excitation; for faster sweeps we observe notably large excursions above the excitation values obtained without monitoring, or even without decay.

LZ transitions subject to monitoring have been studied previously in the context of Landau-Zener-Stueckelberg interferometry \cite{satanin}, with the marked difference of unravelling the dynamics with quantum jumps. If the emitted radiation is detected by photon counting, a single counting event signals the quantum jump, i.e., a transition of the emitter to the ground state from where the dynamics proceeds. Unlike the quantum jumps, which always bring the emitter into its ground state, the continuous random backaction of homodyne detection can lead to both a decrease and an increase in the excited state population \cite{bolund}, and, in conjunction with the LZ Hamitonian terms, to rich dynamics of the emitter.

The article is organized as follows: In Sec. \ref{Model}, we present the dissipative LZ model and the stochastic evolution equations describing the system monitored with a homodyne detection scheme. In Sec. \ref{Simulation} we present numerical simulations and an overview of the main results of the simulations. Sec. \ref{Discussion} closes with a summary and discussion of the results.

\section{Dynamics of the dissipative, probed Landau-Zener model} \label{Model}

\begin{figure}[h]
\centering
\includegraphics[width=0.48\textwidth]{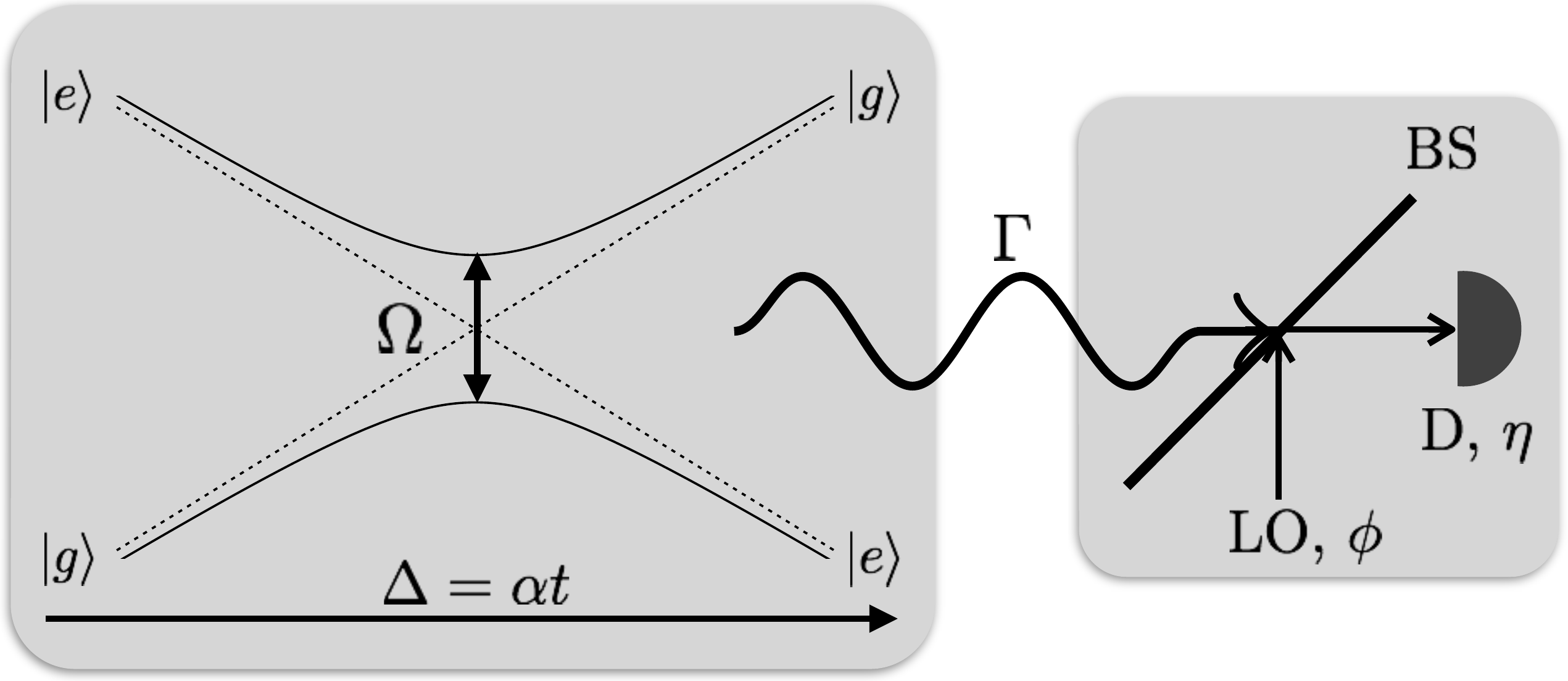}
\caption{The left hand side of the figure depicts the implementation of the Landau-Zener model by the driving of a two-level system between states $|g\rangle$ and $|e\rangle$ with a field of constant amplitude $\Omega$, and with a detuning $\Delta$ that varies linearly with time. The dotted lines show the level crossing with $\Omega=0$, and the solid lines show the adiabatic, time-dependent eigenenergies of the field dressed states. Spontaneous decay from the excited to the ground state with rate $\Gamma$ causes emission of radiation, and in the right hand side of the figure we illustrate the mixing of the signal on a beam splitter (BS) with a local oscillator (LO) with phase $\phi$. The resulting signal is detected in a photodetector (D) with efficiency $\eta$. \label{LZsetup}}
\end{figure}

A two-level system interacting with a near resonant field is governed by the Hamiltonian
\eq \label{LZ}
H = \frac{\hbar\Omega}{2}(\spl+\sm) -\hbar \Delta(t)\spl\sm,
\eeq
where the strength of the coupling is quantified by the Rabi frequency $\Omega$ and $\Delta = \omega_F-\omega_T$ denotes the detuning between the driving field and the two-level transition frequency. This detuning can be made time-dependent by a chirp of the field, or by a modulation of the system eigenenergies, and following the original formulation of the LZ transition problem, we assume a linear chirp such that $\Delta(t)=\alpha\, t$. The lowering and raising (Pauli) operators of the two-level system are $\sigma_-=\ket{g}\bra{e}$ and $\sigma_+=\sigma_-^\dg$.

\begin{figure}[h]
\centering
\includegraphics[width=0.49\textwidth]{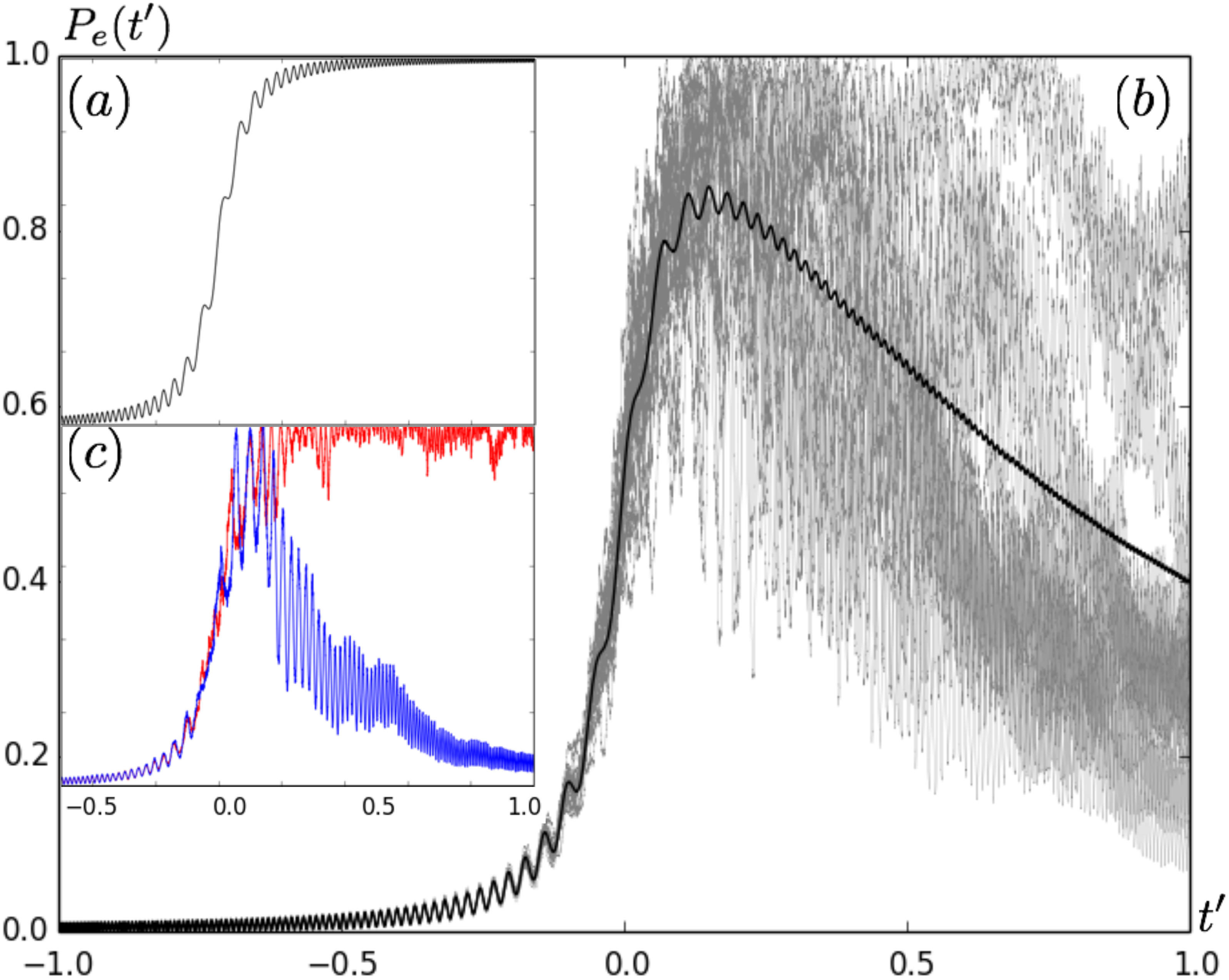}
\caption{Excited state probabilites $P_e(t')$ as a function of $t' = \Gamma t$ for $\alpha=10^4\,\Gamma^2$, $ \Omega=100\,\Gamma$ and $\phi=0$. (a) Solution of the unitary LZ model; (b) A sample of individual trajectories (grey lines) and the average value over 6000 trajectories (black line); (c) Two individual trajectories with exceptionally high (red curve) and low (blue curve) final excited state probabilities. \label{trajectories}}
\end{figure}

We take the initial condition to be $P_e(t_i)=0$, where $P_e(t)=\langle e|\rho|e\rangle$ is the excited state probability. If the laser is chirped infinitely slowly from initial time $t_i=-\infty$, the adiabatic theorem states that at final time $t_f=\infty$ the system becomes excited with unit probability. During a sweep at a finite rate, the asymptotic excited state population is given by the Landau-Zener formula
\eq
P_{LZ}\equiv\lim_{t\rightarrow\infty}P_e(t) = 1-e^{-\gamma\pi/2},
\eeq
where the adiabaticity parameter
\eq
\gamma = \Omega^2/\alpha
\eeq
characterises the speed of the LZ sweep: $\gamma\gg1$ describes the limit of slow, near-adiabatic sweeps and the opposite limit $\gamma\ll1$ fast sweeps. Moreover, in practice the coupling is only turned on for a finite time, $|t_i|, |t_f|<\infty$, leading to modifications of the transition probability \cite{garraway}. An example of the dynamics of a unitary finite-time, finite-speed transition during the level crossing is shown in Fig. \ref{trajectories}a.

The probability of successfully exciting the system during the nonadiabatic passage is further altered when spontaneous decay is taken into account \cite{hanngi}. Under the Born-Markov approximation, decay with a rate $\Gamma$ can be incorporated into a Lindblad master equation for the density matrix $\rho$ of the two-level system,
\eq \label{master_uncond}
d\rho = -\frac{i}{\hbar}[H,\rho]dt + \Gamma \mathcal{D}[\sm]\rho\,dt,
\eeq
where the superoperator $\mathcal{D}[\sigma]$ acts on the density matrix as specified by
\eq
\mathcal{D}[\sigma]\rho = -\frac{1}{2}\left\{\sigma^\dg\sigma,\rho\right\} + \sigma\rho\sigma^\dg.
\eeq
Spontaneous decay typically leads to an exponential reduction in the excited state population and ground-excited state coherence, and the feeding of the ground state associated with the decay. Combined with the chirped coherent excitation of the system, the dynamics of $P_e(t)$ is more involved but we see the main characteristics in the solid black curve of Fig. \ref{trajectories}b: Compared to the unitary evolution, the system is less excited when it reaches resonance, and on a time-scale $t\sim\Gamma^{-1}$, the excitation probability suffers from a roughly exponential decay.

In this article we focus on a homodyne measurement scheme where the atomic radiation signal is mixed with a local oscillator using a highly reflecting beam splitter, and the subsequent signal is measured with detector efficiency $\eta \in [0,1]$. The local oscillator is a strong coherent field characterised by the oscillator phase $\phi$ and depending on the choice of the phase, this measurement scheme provides information on different quadratures $\spl\,e^{-i\phi}+\sm\,e^{i\phi}$ of the atomic dipole via the (suitably normalized) measurement current \cite{jacobs-steck, steck}
\eq \label{current}
dq(t) = \Gamma\sqrt{\eta}\langle\spl\,e^{-i\phi}+\sm\,e^{i\phi}\rangle_\rho dt + \sqrt{\frac{\Gamma}{\eta}} dW.
\eeq
Here $\langle\cdot\rangle_{\rho}$ denotes average with respect to $\rho$. The intrinsic shot noise fluctuations in photon counting are encapsulated in the Wiener increments $dW$. These are Gaussian distributed random variables with zero mean and variance $\text{var}(dW) = dt$. As the field is entangled with the state of the emitter, the measurement of a particular value of the photocurrent $dq(t)$ causes a backaction on the state of the emitter, which is described by the last terms in the conditional, stochastic master equation (SME):
\eq \label{master}
d\rho = -\frac{i}{\hbar}[H,\rho]dt + \Gamma \mathcal{D}[\sm]\rho\,dt + \sqrt{\eta\Gamma}\mathcal{H}[\sm\,e^{i\phi}]\rho\, dW,
\eeq
where the nonlinear homodyne superoperator $\mathcal{H}[\sigma]$ is given by
\eq
\mathcal{H}[\sigma]\rho = \sigma\rho + \rho\sigma^\dg-\langle\sigma+\sigma^\dg\rangle_{\rho}\rho.
\eeq
The random increment $dW$ in (\ref{master}) is the same as the one appearing in the measurement signal (\ref{current}) where, in the modelling of an actual experiment, $dW$ is isolated  by subtracting the expectation value from the detected current. The solution of the SME accompanying the stochastic measurement signal is called a quantum trajectory \cite{carmichael}. We note that a vanishing detector efficiency $\eta=0$ is equivalent to no detection at all, recovering the usual master equation (\ref{master_uncond}).

Figure \ref{trajectories}b shows the excited state population for several quantum trajectories, obtained with $\eta=1$. The average excitation probability over many follows the unobserved density matrix result; the mean for 6000 unravellings is visually indistinguishable from the solution of (\ref{master_uncond}). Mathematically this follows when the property $\langle\langle dW\rangle\rangle=0$ is applied to the SME, where $\langle\langle \cdot\rangle\rangle$ denotes the average over all stochastic realisations. We observe large fluctuations about the average, and despite the decay process, many trajectories seem to reach near unit excitation probabilities at different points in time. Remarkably, among the solutions of the SME it is possible to find trajectories whose dynamics very closely resemble the solution of the undamped LZ problem. One such trajectory is shown is Fig. \ref{trajectories}c as a red curve. In this case detection of the emitted radiation results in a measurement backaction that almost perfectly cancels the effect of decay on the emitter. In the next Section we will elaborate on the properties of such trajectories in more detail.

\begin{figure}[h]
\centering
\includegraphics[width=0.45\textwidth]{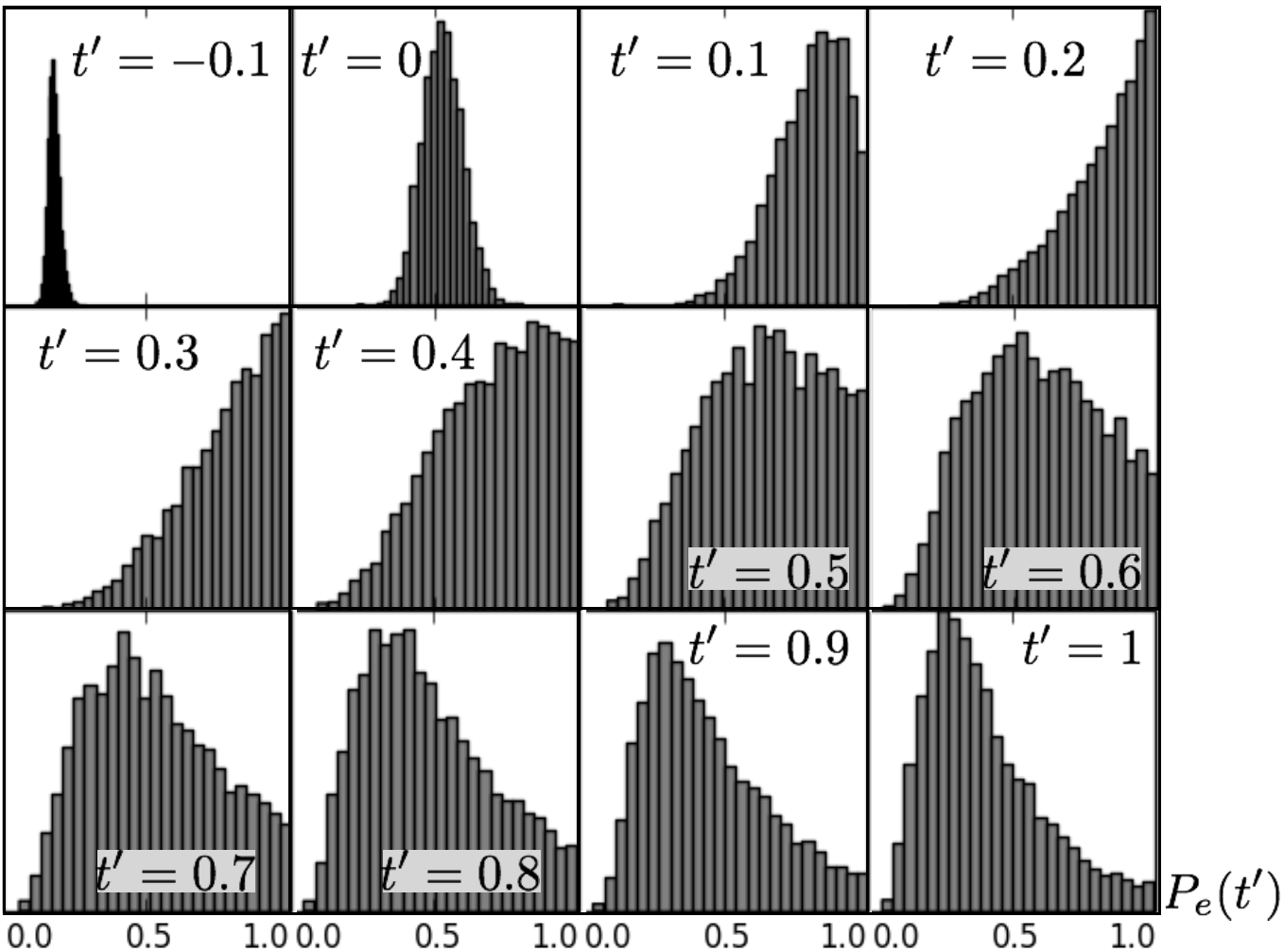}
\caption{\label{histograms}Histograms showing the distribution of the values of $P_e(t')$ for the simulated sample of trajectories at different times $t'=\Gamma t$. The model parameters are the same as in Fig. 2.}
\end{figure}


\section{Results of numerical simulations} \label{Simulation}

We solve the SME (\ref{master}) numerically using the Milstein method, which has strong convergence of $\mathcal{O}(\Delta t)$ \cite{platen}. The data in Fig. \ref{trajectories} is simulated with a time-step of $\Delta t = 4\times10^{-5}\Gamma^{-1}$ in the interval $t'=\Gamma t\in[-1,1]$, and the convergence of the solution has been checked using the method of consistent Brownian paths \cite{steck}. As a first step we describe the general dynamical features of this model for parameter values $\Omega= 100\, \Gamma$ and $ \alpha=10^4\,\Gamma^2 $, corresponding to a slow LZ sweep with $\gamma=10$, and assume perfect detection with $\eta=1$ and $\phi=0$. The effects of inefficient measurements and speeding up the LZ sweep are discussed in later sections.

\subsection{Statistics of quantum trajectories}

The grey lines in Fig. \ref{trajectories}b exemplify the stochastic LZ dynamics and show that the spread about the mean value (of 6000 trajectories) increases after the crossing of the resonance at $t'\approx0$. Intuitively, as long as the occupation probability $P_e(t')$ is small, photons arriving at the detector more likely originate from the local oscillator than from the two-level emitter. The measurement backaction is therefore almost negligible and the individual unravellings do not stray far from the mean. To better quantify the statistics of the trajectories, we use the simulation data to approximate the probability distributions of trajectories at different times with histograms of the excited state population with 25 equal width bins. These histograms are shown in Fig. \ref{histograms}. In the beginning of the Landau-Zener sweep the spread of the trajectories is small, but already after $t'\approx 0.3$ the sample of trajectories effectively cover the whole range of values of $P_e(t')\in[0,1]$.

Interestingly, backaction from the homodyne measurement can drive the values of $P_e(t')$ to much higher values than in the case when the emitted photons are not observed. For $0\lesssim t'\lesssim0.5$ the distributions of the simulated trajectories are characterised by a negative skew\footnote{We have calculated the skew both as the third standardised moment of the distribution and as the Pearson's coefficient.}, implying that the probability density is concentrated to the right, i.e., around high values of the excitation probability. The mode of the histograms is close to unit excitation around time $t'\approx0.2-0.3$ and even as it moves towards lower values, a significant fraction of the trajectories display high values of the excitation probability.

\begin{figure}[h]
\centering
\includegraphics[width=0.4\textwidth]{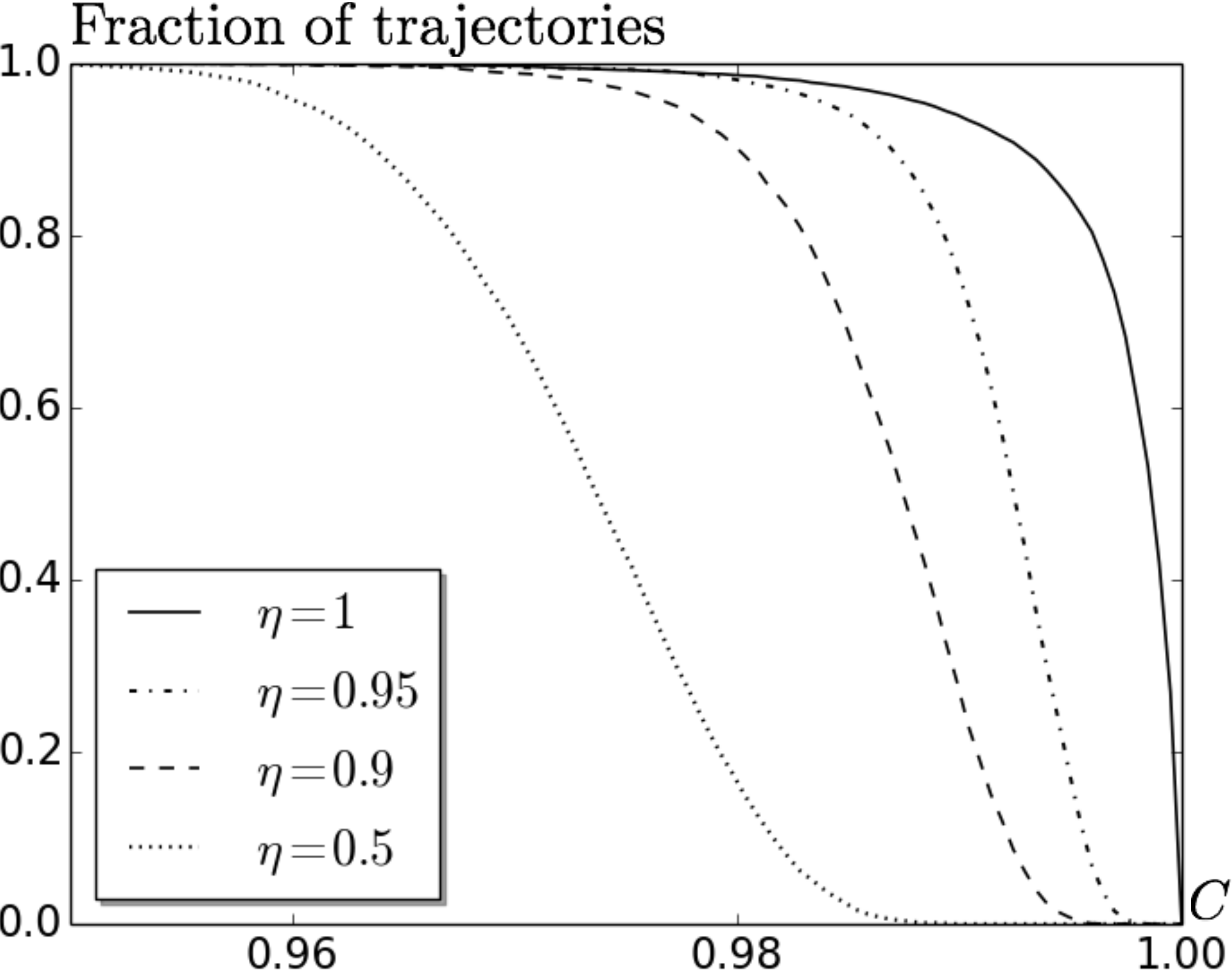}
\caption{\label{exits} The fraction of trajectories that exceed excitation probability $C$ at some point during their evolution for different values of the detector efficiency $\eta$. The model parameters are the same as in Fig. 2.}
\end{figure}

Moreover, our simulations show that it is a \emph{typical} feature, rather than a peculiar property of a small subset of the simulated unravellings of the SME (\ref{master}), to reach high excitation probabilities. Out of the 6000 trajectories simulated with the specified parameters in this Section, every single one reaches a value $P_e(t')=0.96$ for some $t'\in[-1,1]$, while the maximal excitation probability without homodyne detection is about 0.85. The fraction of trajectories with values $P_e(t')\geq\ C$ decreases as $C \rightarrow 1$, but notably slowly: over 94\% of the simulated trajectories reach an excited state probability of 0.99 at some point during the Landu-Zener sweep (See Fig. \ref{exits}, solid line). The blue curve in Fig. \ref{trajectories}c was chosen as representative of a trajectory with a low final value $P_e(t_f)$, but, evidently, also this trajectory reaches almost unit excitation probability at some point during the sweep.

\begin{figure}[h]
\centering
\includegraphics[width=0.4\textwidth]{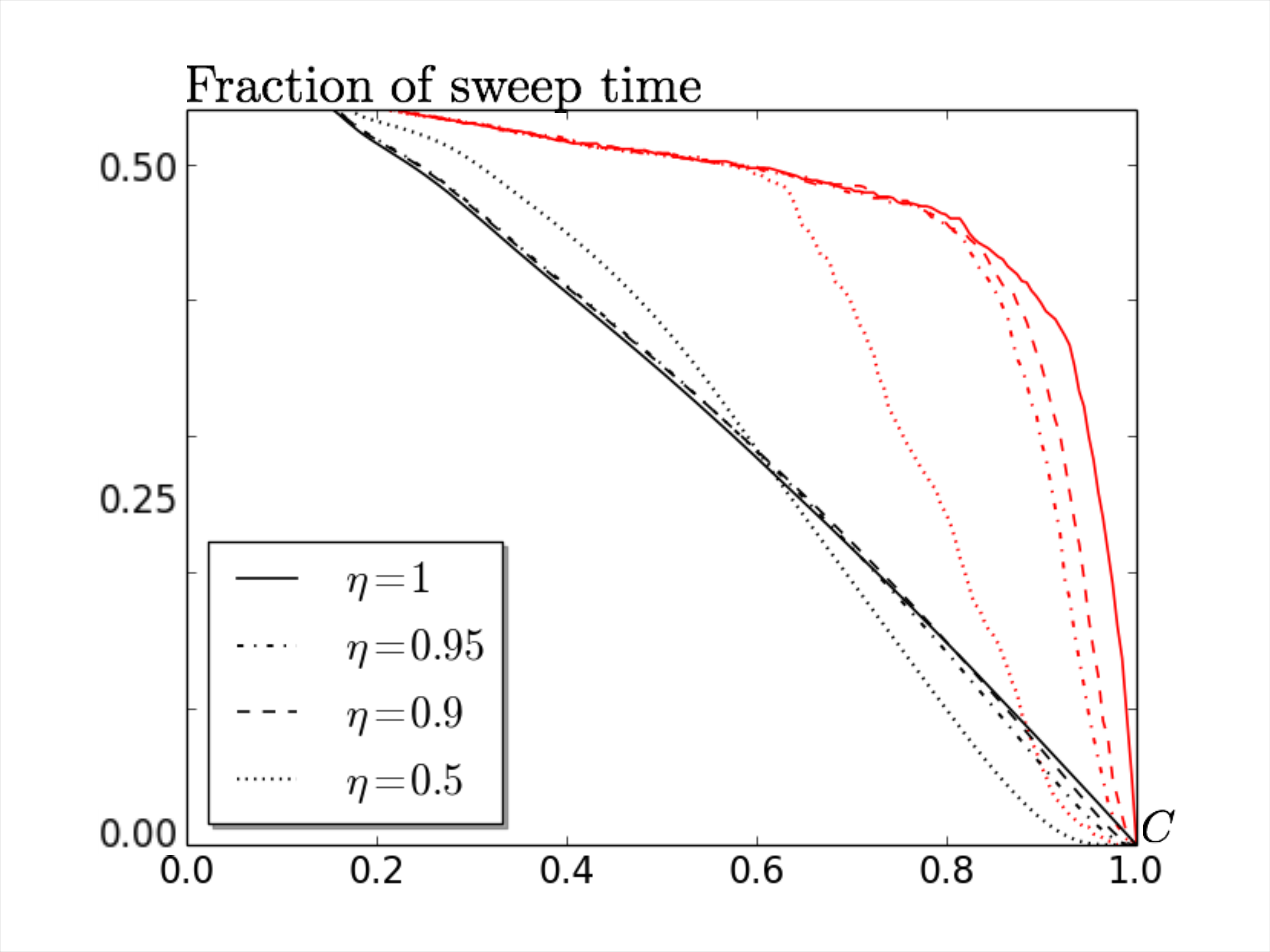}
\caption{\label{times} The maximal (upper, red lines) and average (lower, black lines) fraction of the total sweep time $t'=\Gamma\tau\in[-1,1]$ spent by quantum trajectories with excitation probabilities $P_e(t')\geq C$ for different values of the detector efficiency $\eta$. The model parameters are the same as in Fig. 2.}
\end{figure}

It is natural to ask for how long the high excitation values persist, i.e., given a threshold value $C$, for what fraction of the total sweep time $t'\in[-1,1]$ a trajectory takes values $P_e(t')\geq C$. The mean value of this time, for the sample of simulated trajectories, is shown in Fig. \ref{times}, and it falls approximately linearly to zero with increasing $C$. We also show, with the upper red curve in Fig. \ref{times}, the largest fraction of time obtained within our sample of 6000 trajectories. The maximum value decreases much more slowly than the mean value as a function of $C$. Moreover, the deviations about the mean are large, and we are able to single out trajectories that are able to retain excitations values up to 10 longer than the average. One such trajectory is displayed as the red curve in Fig. \ref{trajectories}c.

In Ref. \cite{bolund}, it was shown that in the absence of an exciting field, the probability for a decaying two-level system to become conditionally excited during homodyne detection depends crucially on the local oscillator phase. In our case, however, adopting different values for the phase of the local oscillator field turns out to have negligible effect on the main features of the quantum trajectories. We ascribe this to the rapidly rotating frame induced by the detuning sweep, and we imagine that a corresponding sweep of the local oscillator phase may be needed to reveal an effect of the phase on the excitation dynamics.

\begin{figure}[h]
\centering
\includegraphics[width=0.4\textwidth]{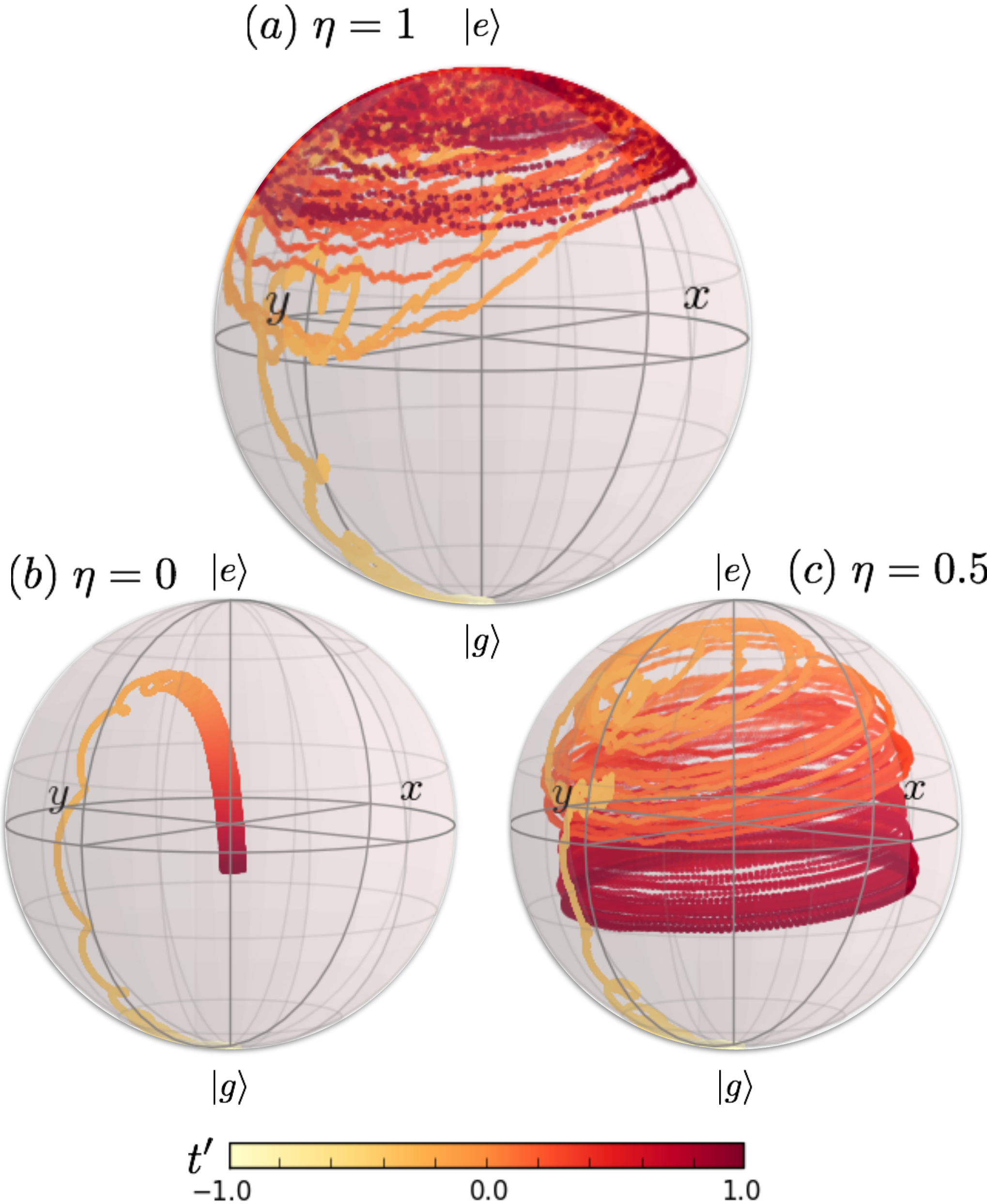}
\caption{\label{balls} Bloch sphere picture of individual quantum trajectories according to the stochastic master equation (\ref{master}). Results are shown with the same parameters as in Fig. \ref{trajectories} for three different values of the detection efficiency $\eta$. In (a) $\eta=1$ and the trajectory stays on the surface of the Bloch sphere. In (b) $\eta=0$, corresponding to trajectory shown in the black line in Fig. \ref{trajectories}b. In (c) $\eta=0.5$ and as in part (a) of the figure, we observe a strong angular precession and diffusion, but with a shorter Bloch vector, thus preventing perfect excitation of the emitter.}
\end{figure}

\subsection{Modifications due to inefficient detection \label{inefficient}}

Homodyne detection plays a pivotal role in the ability of the two-level system to reach high excitation values during the noisy LZ sweep. Detector efficiency $\eta$ is an important figure and with lower $\eta$ we expect a shift in the probability distributions of the trajectories towards lower values of $P_e(t')$, as well as a suppression of the fraction of rare trajectories with very high values of the excitation probability. Quantitatively, the dependence of the results on the detector efficiency is shown in Fig. \ref{exits}. Clearly the ability for a large fraction of the simulated frequencies to reach a high excitation value is drastically reduced. For example, with $\eta=0.95$ only 75\% of the simulated trajectories reach probability excitation value of 0.99 at some point during the LZ sweep, and for $\eta=0.9$ this fraction drops to about 27\%, while of the simulated trajectories for $\eta=0.5$ none reach value $P_e(t')=0.99$. The average and maximal times the trajectories spend above a given value of $C$ are less effected for moderate reductions of the the detection efficiency, as shown in the insert of Fig. \ref{times}, and it it still possible to find trajectories that retain relatively high excitation values for long times.

The reason for the reduced ability of individual trajectories to reach high excitation values is captured in Fig. \ref{balls}, which shows individual trajectories, mapped on the Bloch sphere\footnote{The figures were made using the QuTiP package \cite{qutip}.}, for different detection efficiencies. Perfect detection with $\eta=1$ ensures maximal knowledge of the photoemission record obtained during each simulated run of the experiment, which in turn leads to a pure state of the two-level system throughout the LZ sweep. The Bloch vector representing the system under perfect field detection retains its unit length and the random kicks from the measurement backaction cause diffusive motion on the surface of the Bloch sphere. Together with the rapid precession of the Bloch vector due to $\Omega$ and $\Delta$, this diffusion eventually drives the system close to the north pole of the Bloch sphere, representing the excited state of the emitter. With lower detector efficiencies a fraction of the emitted radiation remains unobserved, leading to a mixed state of the emitter, i.e., a Bloch vector with less than unit length.  The less efficient the detection, the further the trajectories stray from the surface of the Bloch sphere, and while the Bloch vector may acquire a direction towards the north pole due to the angular diffusion and the precession, its shorter length directly leads to the reduction in the excited state population.

\begin{figure}[h]
\centering
\includegraphics[width=0.41\textwidth]{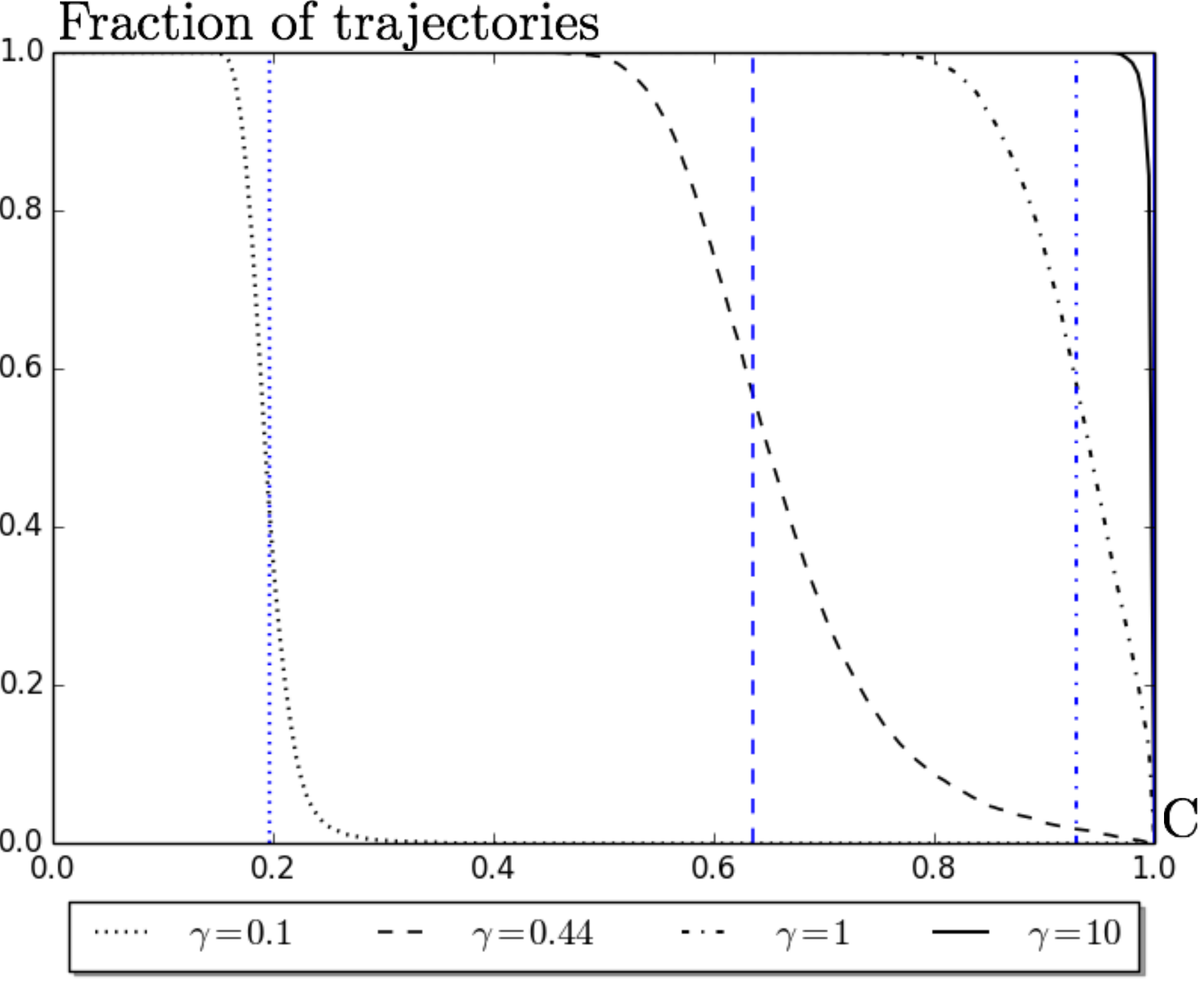}
\caption{\label{fast} Fraction of trajectories (black lines) that exceed excitation probability $C$ at some point during their evolution for different values of the adiabaticity parameter $\gamma$. The blue vertical lines represent the maximum value during a unitary LZ sweep for the respective values of $\gamma$. All parameter values are the same as in Fig. \ref{trajectories}, except for the Rabi frequency $\Omega$, which has been modified to correspond to each value of $\gamma$.}
\end{figure}

\begin{figure}[h]
\centering
\includegraphics[width=0.41\textwidth]{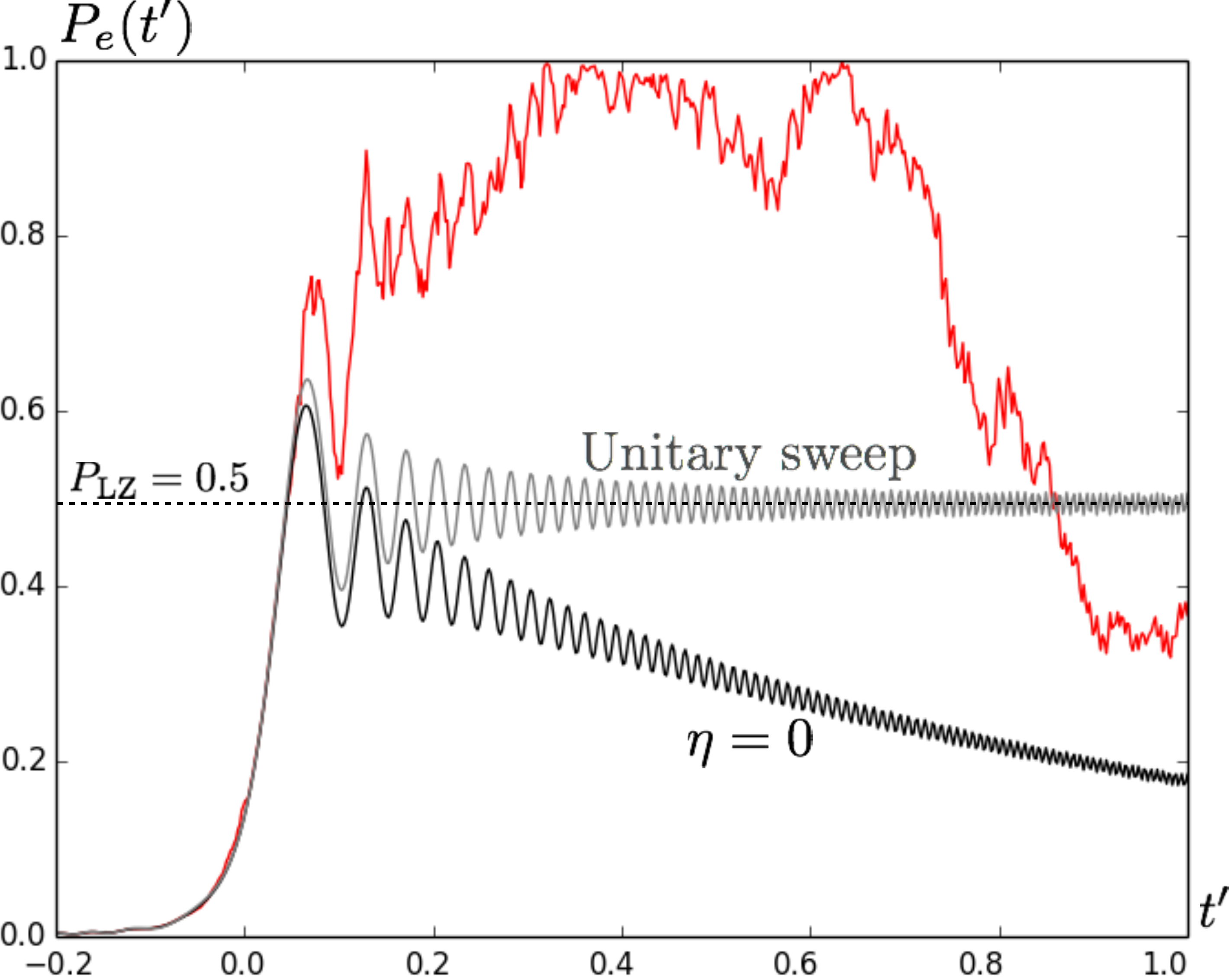}
\caption{\label{fast2} The excitation probability $P_e(t')$ for a single trajectory (top red curve), for unconditioned dynamics (bottom black curve) and for a unitary LZ sweep (middle grey curve). The parameter values are the same as for the intermediate speed sweep with $\gamma=0.44$ in Fig. \ref{fast}.}
\end{figure}

\subsection{Speeding up the LZ sweeps \label{speed}}

In the simulations presented so far the LZ sweep has been  near-adiabatic with a transition probability $1-P_\text{LZ}\approx1.5\times10^{-7}$ in the absence of noise. The two-level system attains high excitations values soon after crossing the resonance, leading to high probability of spontaneous emission and greater rate of detected photons arriving from the emitter. Consequently we found that the measurement backaction can significantly alter the dynamics, especially with high detection efficiencies. With fast sweeps the value of $P_e(t')$ remains low throughout the LZ sweep, the effect of backaction is reduced and the trajectories do not deviate much from the mean. This is reflected in decreased ability of the quantum trajectories to reach high excitation values, as shown in Fig. \ref{fast}. For a sweep with $\gamma=0.1$, for example, none of the 6000 simulated trajectories reach a value higher than 0.4.

For moderate sweep speeds the backaction remains sufficiently strong to almost fully excite the two-level system. In Fig. \ref{fast2} we show one such trajectory for a sweep speed $\gamma \approx 0.44$, corresponding to half-excitation of a unitary LZ sweep, $P_\text{LZ}=1/2$. Note that now the diffusion, which is associated with a monitoring of the decay process, causes the trajectory to reach excitation probabilities which are considerably higher than if the system had evolved in a unitary manner with a vanishing excited state decay rate $\Gamma$. This phenomenon is somewhat typical with about 56\% of the simulated trajectories exceeding value $P_U^\text{max}=0. 64$, where $P_U^\text{max}$ is the maximal excitation value during a unitary LZ sweep. The vertical lines in Fig. \ref{fast} decipher $P_U^\text{max}$ for each respective value of the adiabaticity parameter $\gamma$ and we conclude that a fair fraction of the trajectories become more excited when they decay, in comparison to the unitary LZ sweeps, irrespective of the sweep speed.

\section{Conclusions} \label{Discussion}

We have studied a two-level quantum system driven across an avoided crossing when the radiative emission from the system is monitored with a homodyne measurement. The measurement backaction results in diffusive dynamics and enables considerable deviations from the dynamics followed by the system in absence of monitoring. When the emission is detected with unit efficiency, the measurement preserves the purity of the decaying system and confines all quantum trajectories to the Bloch sphere surfacce. The random diffusion allows the system to explore the entirety of the Bloch sphere and obtain, for example, excitation levels higher than one would observe in the absence of monitoring, and even in the absence of radiative decay. A growing number of physical systems now permit efficient continuous monitoring, and they may serve to demonstrate and to benefit from measurement backaction dynamics, such as the result obtained in this article. The large excursions from the ensemble mean values observed here are likely to apply also to other physical properties, including the emergence of squeezing and entanglement due to measurement backaction on both few particles \cite{sm}, and on many particle systems \cite{brian}.

\begin{acknowledgments}
We thank S\o ren Gammelmark for useful discussions and acknowledge funding from the European Commission (ITN CCQED) and the Villum Foundation.
\end{acknowledgments}

\end{document}